# NFT Hydroponic Control Using Mamdani Fuzzy Inference System

Indra Agustian [1*], Bagus Imam Prayoga [2], Hendy Santosa [3], Novalio Daratha [4], Ruvita Faurina [5]
[1, 2, 3, 4] Department of Electrical Engineering, Universitas Bengkulu, Indonesia
[5] Department of Informatics, Universitas Bengkulu, Indonesia
Email: [1] indraagustian@unib.ac.id, [2] bagusimamprayoga@gmail.com, [3] hendysantosa@unib.ac.id, [4] ndaratha@unib.ac.id,
[5] ruvita.faurina@unib.ac.id
*Corresponding Author

*Abstract*—The Nutrient Film Technique (NFT) method is one of the most popular hydroponic cultivation methods. This method has advantages such as easier maintenance, faster and optimal plant growth, better use of fertilizers, and less deposition. The disadvantages of NFT include the consumption of electrical power and the faster spread of disease. Therefore, NFT requires a good nutrient control and monitoring system to save electricity and achieve optimal growth and resistance to pests and diseases. In this study, a nutrient control was designed with indicators of pH and TDS levels and equipped with an Internet of Things (IoT) based monitoring system. The control system used is the Mamdani Fuzzy Inference System. The output of the system is the active time of the pH Up, pH Down, and AB Mix nutrient pumps, which aim to normalize the pH and TDS of nutrient liquids. The experimental results show that one to three control steps are needed to normalize pH. One control step has a response time of 60 seconds, and it can prevent pH Up and pH Down oscillations. As for TDS control, the prediction of AB mix pump active time works accurately, and TDS levels can be normalized in one control step. Overall, based on surface control, simulations, and real experimental data, it is indicated that the control system operates very well and can normalize pH and TDS to the desired normal standard.

I. INTRODUCTION

The collaboration of automation technology and cyber technology has caused major changes in all aspects of human life in a short time. In the industrial sector, these changes are known as the industrial revolution 4.0 [1] [2] [3]. The collaboration makes various systems in the industrial world easier to manage, more effective, and more efficient, thus becoming more productive [4] [5].

One of the fields that can take advantage of the collaboration of automation technology is the hydroponic cultivation field [6] [7]. Generally, hydroponic cultivation is done conventionally, with manual system operation, and some of them have implemented simple automation technology with a timer-based automation system. There are still a few who develop smart and precise farming systems that collaborate with automation technology and cyber technologies such as the Internet of Things (IoT) [8] [9] [10] [11]. Therefore, there are still many research and development topics that can be done to improve the quality of hydroponic cultivation.

Hydroponics is a method of cultivating plants by utilizing water without soil planting media and emphasizing the fulfillment of nutrients needed by plants [12] [13]. For certain plants, hydroponic agricultural cultivation can use water more effectively and efficiently than cultivation using soil growing media. This method is especially suitable for areas with limited land and water supply [14].

Nutrient Film Technique (NFT) is a popular hydroponics cultivation system [15] [16]. NFT hydroponics offers advantages, such as easier maintenance, faster growth, better use of fertilizers, and less deposition. However, NFT has drawbacks such as dependence on electrical power, speed of spread of disease, and installation costs [17]. Many studies have been carried out on the control of NFT hydroponic nutrients. The control systems that have been widely developed are the Proportional Integral Derivative (PID) and Fuzzy methods. In addition, there are also those using predictive control methods with the Deep Neural Network (DNN) and K-Nearest Neighbor (KNN) algorithms.

Research [18] designed an NFT hydroponic nutrient control system based on the PID method [19] [20] by entering pH level data while controlling the air level using the ON/OFF method as research [21] did. Research [22] designed an NFT hydroponic nutrient control system using the PID method with inputs in the form of water level and electrical conductivity [23] of the nutrient liquid.

Research [24] designed a nutrient control system on NFT-based hydroponic plants using the multiple linear regression method. The system only automated the pH of hydroponic nutrient. Research [25] designed a hydroponic nutrition control system for lettuce NFT by using the predictive value obtained through the DNN (Deep Neural Network) algorithm [26] [27]. This research uses 4000 nutrition control data as input of a DNN. Research [28] designed an NFT hydroponic nutrient control system based on demand prediction with the KNN algorithm [29] [30] [31], with input pH, TDS, electrical conductivity, and temperature. The predicted results are the ON/OFF state of the pH and nutrient pumps.

The following explanations are previous studies that used the fuzzy method to control nutrients in hydroponics. Research [32] conducted the control of Electrical Conductivity (EC) [33] [34] of hydroponics nutrients based on volume and the error of EC to setpoint using a fuzzy controller, The output is the active duration of the nutrient pump. Research [35] conducted a smart-greenhouse control simulation using a fuzzy logic controller. The control output is the duration of watering based on temperature and humidity. Research [36] controlled the electrical conductivity and pH in hydroponics using the fuzzy logic method, the output is a combination of three output states (ON/OFF) with a fixed duration. Research [37] conducted EC control using fuzzy logic based on the HSV image processing technique [38] on the Pakcoy mustard plant. The input is the EC error to the setpoint and water level. The output is the active duration of the nutrient and water pump in the same fuzzy set. Research [39] adjusted the duration of activation of water pumps and nutrient pumps based on EC and NFT water levels using the Mamdani FIS. Research [40] conducted a simulation to regulate the temperature, humidity, EC, and pH of nutrients in greenhouse hydroponics using FIS Sugeno. Research [41] controlled the pH and EC hydroponics using fuzzy with the output of the active time duration of the pump pH Up and pH Down. Research [42] controlled the pH and humidity of hydroponics using fuzzy with the output of the active duration of the pH pump and fan. Research [43] controlled EC, pH, and water level using fuzzy logic in hydroponic NFT, with the output of the active duration of the water pump, nutrient concentrate tank pump, and nutrient tank drain. Research [44] controlled nutrient concentrations in hydroponics using FIS Sugeno, with PPM error to setpoint and delta error as the inputs, and the active duration of nutrient and water pumps as outputs.

Research [45] controlled the EC and pH of hydroponic NFT using fuzzy. With input EC, pH temperature, light intensity, and humidity, the output is the active duration of the water pump, nutrient solution, acid solution, and the alkaline solution. Research [46] controlled EC and pH using fuzzy logic in hydroponic NFT with input EC and pH, resulting output the active duration of the pump of EC solution, water, acid solution, and the alkaline solution. Research [47] [48] is almost the same, adjusting the pH of hydroponic NFT using fuzzy logic with the output of the active duration of the acid and water pump.

From the hydroponic fuzzy controller studies mentioned above, based on the type of inference, there are two types of FIS used, FIS Mamdani [49] and FIS Sugeno [50]. Some of the studies above did not mention and explain the FIS carried out, but based on the fuzzy rules they created, the FIS they used was the Mamdani FIS. Meanwhile, those who use FIS Sugeno explicitly mention the FIS. Overall, the control carried out commonly is on EC and pH levels, and the control output is the active duration of certain pumps.

FIS Sugeno works better for linear control such as PID, optimization, and adaptive techniques, but FIS Sugeno is well-suited for MISO (Multiple Input and Single Output) systems and systems that require mathematical analysis [50]. While FIS Mamdani can be used for MIMO (Multiple Input and Multiple Output) systems [51], it is well-suited for applications with rules created based on human expert knowledge. FIS Sugeno has the advantage of being computationally more efficient than FIS Mamdani, but FIS Mamdani is more intuitive with the output of each rule being a fuzzy set [52].

Compared to the ON/OFF method, the PID, FIS Sugeno, and FIS Mamdani control systems are much more effective. However, Sugeno's PID and FIS are well-suited for the MISO system, while the MIMO system is well-suited for the Mamdani FIS [53]. In this study, the controlled system is a MIMO system, not linear, which does not require mathematical control but is more intuitive, so FIS Mamdani is the most suitable control choice.

In this study, a closed-loop control system carried out the Mamdani Fuzzy Inference System (FIS-Mamdani) method to control the pH and TDS levels of NFT hydroponic nutrients. The closed-loop control system can reduce electricity consumption compared to the timer-based control. And also can control nutrient quality better to increase plant resistance to disease. The inputs are pH and TDS levels, and the output is the active duration of the pH Up pump, pH Down pump, and nutrient solution pump. The previous study that controlled concentrations based on TDS is research [44], but this study used the FIS Sugeno controller. In other previous studies, the nutrient concentration were commonly determined based on EC.

Based on the explanation above, the contribution of this research is the implementation strategy of the FIS Mamdani controller on the MIMO system of NFT Hydroponics. Using a microcontroller as CPU to adjust the active time duration of the pH Up pump actuator, pH Down pump, and nutrient solution pump based on data from pH and TDS sensors, to control the required pH and TDS level.

Furthermore, in this article, section 2 describes the control hardware and the Mamdani FIS control system design. Section 3 discusses the results of the design and performance of the control system in regulating the pH and TDS levels of NFT hydroponic nutrient solution. And section 4 explains the conclusions of the research.

## II. MATERIAL AND METHODS

### A. NFT Hydroponic Design

In this study, the NFT hydroponic rack was designed using 4-inch PVC pipes arranged vertically as shown in Fig. 1. The rack dimensions are 150 cm long, 50 cm wide, and 250 cm high. The pipes are tilted 5 degrees so the water can flow at a low intensity. The nutrient solution is drawn from the reservoir by a 12 volt DC pump with a flow rate of 1 liter per minute, then connected to a 3/16 inch hose into the hydroponic inlet pipe, and the hydroponic outlet pipe is connected to the 1/4 inch return pipe to the nutrient reservoir.

In the nutrition system, there are two storage media, a bucket with a capacity of 25 liters as a nutrient solution reservoir and three bottles with a size of 1 liter as a storage medium for concentrated liquid pH Up, pH Down, and fertilizer. The pH concentrate and fertilizer are drawn by a 12 volt DC diaphragm water pump with a flow rate of 3.3 liters per minute, which is activated using a relay contactor.

Fig. 2 shows the main electronic devices and the connection directions of the system. Fig. 3 shows the electronic circuit of the control system and IoT monitoring. Details of the electronic devices supporting the NFT hydroponic nutrition control system and IoT in this study are shown in Table 1. The sensors used are TDS sensors SEN0244 [54], pH sensors SEN0161 [55], and ultrasonic sensors [56]. Those sensor data are the system input in the form of the number of dissolved particles in the water, the pH level, and the water level. The control computation uses the Arduino Mega 2560 Microcontroller [57] which provides instructions to activate the actuator pump for pH, nutrients, and water solutions. The designed NFT hydroponics is also equipped with an IoT monitoring system to monitor the values of each parameter. IoT system uses the ESP01 IoT Module [58] and the Blynk application framework as user interface [59].

The flow of nutrient liquid that comes out and the addition of fertilizer can change the pH and TDS levels of the nutrient liquid. Controlling the pH and TDS of this nutrient liquid is carried out based on data from the pH sensor and TDS sensor and then processed using the Mamdani fuzzy inference method by a microcontroller. The pH and TDS standards used were the kale plant pH and TDS standards with pH 5.5-6.5 and TDS 1050-1400 ppm.

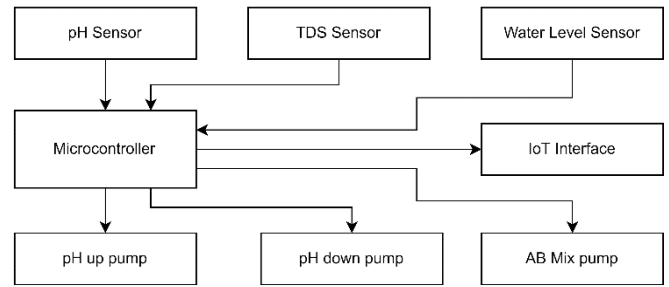

Fig. 2. The main supporting electronic devices for control systems and IoT

### B. Mamdani Fuzzy Inference System Design

The Mamdani fuzzy inference system (FIS Mamdani) was originally introduced as a method for developing a control system by synthesizing a set of linguistic control rules obtained from experienced human operators [60, 61, 62]. In FIS Mamdani, the output of each rule is a fuzzy set. The input of the control system is the numerical values of pH and TDS levels. In the fuzzification process, the numerical values of pH and TDS are mapped into fuzzy sets, and the degree of membership is determined in the fuzzy set. Fig. 4 shows the FIS Mamdani architecture used in this study.

The pH membership set is divided into five: strong acid, weak acid, normal, weak alkaline, and strong alkaline. The range of membership degrees shown in (1) – (5). The determination of the TDS membership set is determined based on pre-study test data, which shows the lowest TDS level on average is 150, and the highest average is 1400 ppm.

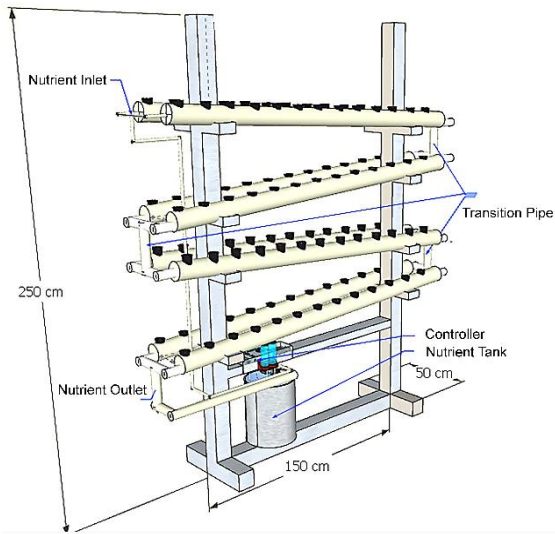

Fig. 1. NFT hydroponic design

Water circulation in NFT hydroponics is the same as circulation in an aquarium. If the water volume is less than 20 liters, the water pump will be active to fill the lack of water volume with a tolerance gap of 1 liter. The water volume is controlled based on the water level in the nutrient reservoir. The water level is measured using an ultrasonic sensor.

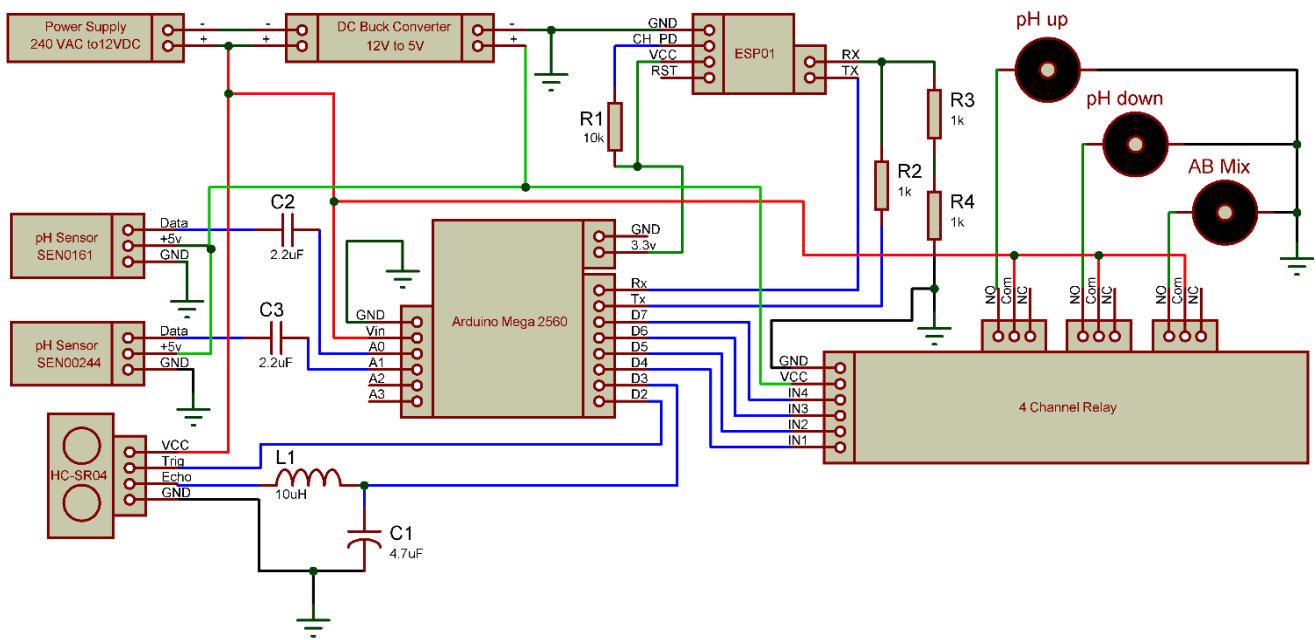

Fig. 3. The electronic circuit of control system and IoT monitoring

By assuming if the TDS is more than 1400 ppm then the addition of fertilizer is stopped, this is intended to simplify control so that the fuzzy set is only divided into three: very low, low and normal. The mapping of the numerical value of TDS to the degree of membership in the fuzzy set is shown in (6) – (8) and visualized using python programming [63] and scikit-fuzzy library [64] in the graph in Fig. 5.

TABLE I. CONTROL SYSTEM ELECTRONIC DEVICES

| No | Device | Qty | Specification |
|---|---|---|---|
| 1 | Microcontroller Arduino Mega 2560 | 1 | *Input Voltage* 7-20V, *Digital I/O* 54 pins, *Analog Input* 16 pins, 16 MHz, 40mA |
| 2 | TDS Sensor SEN0244 | 1 | *Operating Voltage* 3.3 – 5.5V DC, *Measuring Range* 0 – 2000 ppm |
| 3 | pH Sensor SEN0161 | 1 | *Module Power* 3.3 – 5.5V DC, *Measuring Range* 0 – 14 pH |
| 4 | ESP01 IoT Module | 1 | *Operating Voltage* 3.3V DC, *Power Transmission Range* 100m, CCK 1 MBps, UHF 2.4 Ghz. |
| 5 | Ultrasonic Sensor HC-SR04 | 1 | *Module Power* 5V DC, *Measuring Range* 2 – 400cm |
| 6 | 4 channel relay module | 1 | *Input Voltage* 12V, *Max Contact* 250V AC, 10A & 30V DC, 10A |
| 7 | DC Motor Water Pump | 1 | *Operating Voltage* 12V DC, 360 L/H, 450mA, 5W |
| 8 | DC Motor pH Pump and Nutrient Solution Pump | 3 | *Operating Voltage* 12V DC, 3W, 200 L/H, 250mA |
| 10 | *Power Supply* SMPS | 1 | *Output Voltage* 12V DC, 10A |
| 11 | *Buck Converter* LM2596 | 1 | *Max Input Voltage* 40V, *Output Voltage* 1.2 – 37V DC, ±4% |

The output of FIS-Mamdani is the active time of the actuator pump for pH Up, pH Down, and fertilizer concentrates (AB mix). The fuzzy set outputs are shown in Fig. 6. The determination of the numerical value mapping to the degree of membership in the fuzzy set is based on the fastest and longest time test data as shown in (9) – (14).

$$\mu_{pH-strong\ acid}[x] = \begin{cases} 1; & x \leq 1 \\ \frac{4-x}{4-1}; & 1 < x < 4 \\ 0; & x \geq 4 \end{cases} \quad (1)$$

$$\mu_{pH-weak\ acid}[x] \begin{cases} 0; & x \leq 1\ or\ x > 5.5 \\ \frac{x-1}{4-1}; & 1 < x < 4 \\ \frac{5.5-x}{5.5-4}; & 4 \leq x \leq 5.5 \end{cases} \quad (2)$$

$$\mu_{pH-normal}[x] \begin{cases} 0; & x \leq 4\ or\ x > 8 \\ \frac{x-4}{5.5-4}; & 4 < x \leq 5.5 \\ 1; & 5.5 < x < 6.5 \\ \frac{8-x}{8-6.5}; & 6.5 \leq x \leq 8 \end{cases} \quad (3)$$

$$\mu_{pH-weak\ alkaline}[x] \begin{cases} 0; & x \leq 6.5\ or\ x \geq 11 \\ \frac{x-6.5}{8-6.5}; & 6.5 < x < 8 \\ \frac{11-x}{11-8}; & 8 \leq x \leq 11 \end{cases} \quad (4)$$

$$\mu_{pH-strong\ alkaline}[x] = \begin{cases} 0; & x \leq 8 \\ \frac{x-8}{11-8}; & 8 < x < 12 \\ 1; & x \geq 12 \end{cases} \quad (5)$$

$$\mu_{TDS-very\ low}[x] = \begin{cases} 1; & x \leq 150 \\ \frac{625-x}{625-150}; & 150 < x < 625 \\ 0; & x \geq 625 \end{cases} \quad (6)$$

$$\mu_{TDS-low}[x] = \begin{cases} 0; & x \leq 150\ or\ x \geq 1050 \\ \frac{x-150}{625-150}; & 150 < x < 625 \\ \frac{1050-x}{1050-625}; & 625 \leq x \leq 1050 \end{cases} \quad (7)$$

$$\mu_{TDS-normal}[x] = \begin{cases} 0; & x \leq 625\ or\ x > 1400 \\ \frac{x-625}{1050-625}; & 625 < x < 1050 \\ 1; & 1050 \leq x \leq 1400 \end{cases} \quad (8)$$

$$\mu_{pH\ up-fast}[x] = \begin{cases} 0; & x \leq 0\ or\ x \geq 1800 \\ \frac{1800-x}{1800-300}; & 300 < x < 1800 \\ 1; & x \leq 300 \end{cases} \quad (9)$$

$$\mu_{pH\ up-slow}[x] = \begin{cases} 0; & x \leq 300\ or\ x \geq 3000 \\ \frac{x-300}{1800-300}; & 300 < x < 1800 \\ 1; & x \geq 1800 \end{cases} \quad (10)$$

$$\mu_{pH\ down-fast}[x] = \begin{cases} 0; & x \leq 0\ or\ x \geq 1800 \\ \frac{1800-x}{1800-300}; & 300 < x < 1800 \\ 1; & x \leq 300 \end{cases} \quad (11)$$

$$\mu_{pH\ down-slow}[x] = \begin{cases} 0; & x \leq 300 \\ \frac{x-300}{1800-300}; & 300 < x < 1800 \\ 1; & x \geq 1800 \end{cases} \quad (12)$$

$$\mu_{AB-fast}[x] = \begin{cases} 0; & x \leq 0\ or\ x \geq 2400 \\ \frac{2400-x}{2400-400}; & 400 < x < 2400 \\ 1; & x \leq 400 \end{cases} \quad (13)$$

$$\mu_{AB-slow}[x] = \begin{cases} 0; & x \leq 400 \\ \frac{x-400}{2400-400}; & 400 < x < 2400 \\ 1; & x \geq 2400 \end{cases} \quad (14)$$

After the fuzzification process is complete, the next step is to create the fuzzy rules. The fuzzy rules created are as follows:

1. IF pH is a strong acid AND TDS is very low THEN pH Up pump is slow AND the AB mix pump is slow.
2. IF pH is a strong acid AND TDS is low THEN pH Up pump is slow AND the AB mix pump is fast.
3. IF pH is a strong acid AND TDS is normal THEN pH Up pump is slow.
4. IF pH is weak acid AND TDS is in very low THEN pH Up pump is fast AND AB mix pump is slow.
5. IF pH is weak acid AND TDS is low THEN pH Up pump is fast AND the AB mix pump is fast.
6. IF pH is weak acid AND TDS is normal THEN pH Up pump is fast.

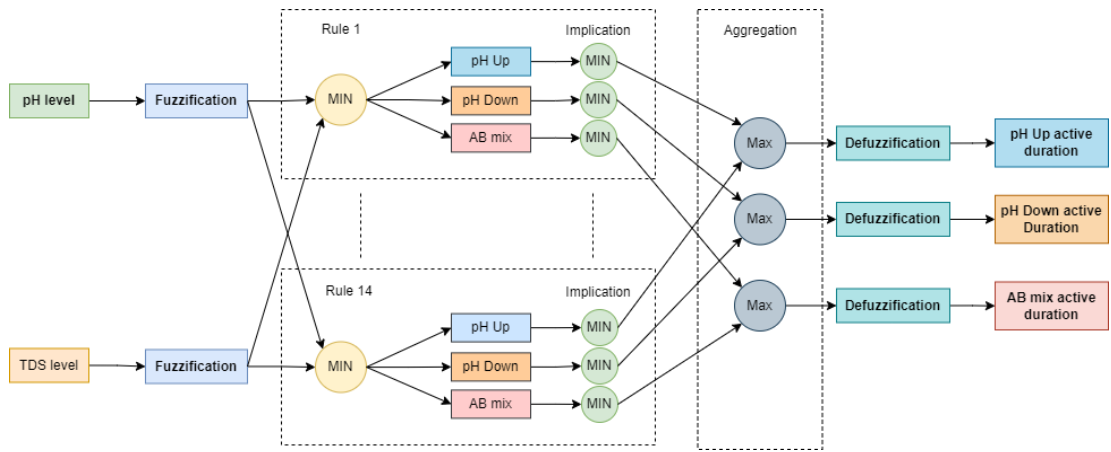

Fig. 4. FIS Mamdani Architecture

7. IF pH is normal AND TDS is very low THEN AB mix pump is slow.
8. IF pH is normal AND TDS is low THEN AB mix pump is fast.
9. IF pH is weak alkaline AND TDS is very low THEN pH Down pump is fast and AB mix pump is slow.
10. IF pH is weak alkaline AND TDS is low THEN pH Down pump is fast AND AB mix pump is fast.
11. IF pH is a weak alkaline AND TDS is normal THEN the pH Down pump is fast.
12. IF pH is a strong alkaline AND TDS is very low THEN the pH Down pump is slow AND the AB mix pump is slow.
13. IF pH is a strong alkaline AND TDS is low THEN the pH Down pump is slow AND AB mix pump is fast.
14. IF pH is a strong alkaline AND TDS is normal THEN the pH Down pump is slow.

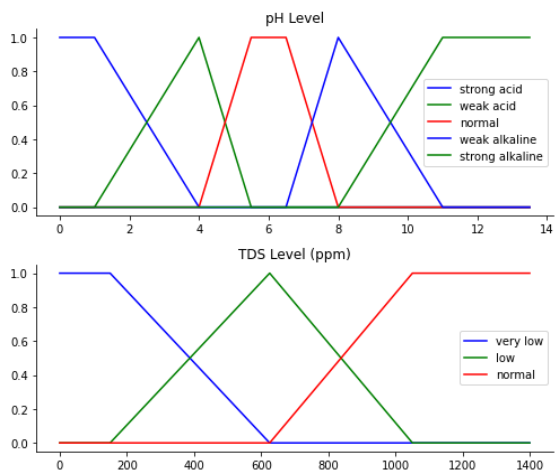

Fig. 5. Fuzzification graph of the input fuzzy set

From the fuzzy rules above, logical operations and implications are then carried out. As shown in Because in the fuzzy rules, the antecedent only uses the logic "AND" then the logical operation uses the min function. The implication function used is also min function. The outputs of the IF THEN rules are combined into a single fuzzy set. The aggregation method used is the max function for all the outputs of the IF THEN rules. The min function on the implications and the max function on the aggregation of the Mamdani method are also called the min-max inference method. The defuzzification method for mapping the magnitude of the fuzzy set into a numeric number is the center of gravity method, which is also known as the centroid method [65] [66].

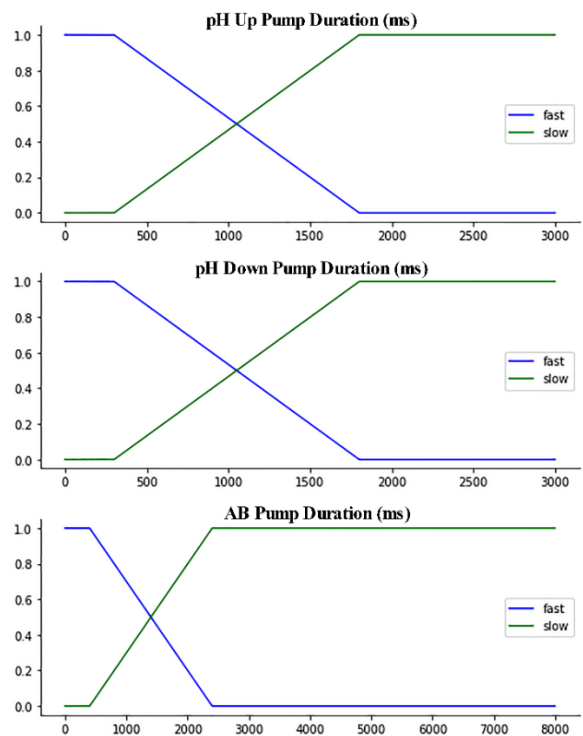

Fig. 6. Fuzzification graph of the output fuzzy set

### III. RESULT AND DISCUSSION

#### A. Control Surface

Before conducting real experiments, first, a simulation is carried out to determine the control surface of each fuzzy output from the FIS Mamdani controller that has been designed. The control surface for pH (Up/Down) control from the simulation results is shown in Fig. 7. In the range 0<pH<5.5 is the control surface of pH Up, and 6.5<pH≤14 is the control surface of pH Down, while for normal pH, the active time is equal to zero. Based on the surface control of both pH Up and pH Down, it can be seen that the pH control design with the Mamdani FIS does not experience significant anomalies and can work well. Likewise, the AB mix surface control as shown in Fig. 8, can be seen that the performance of the AB mix control also does not experience significant anomalies and can work well.

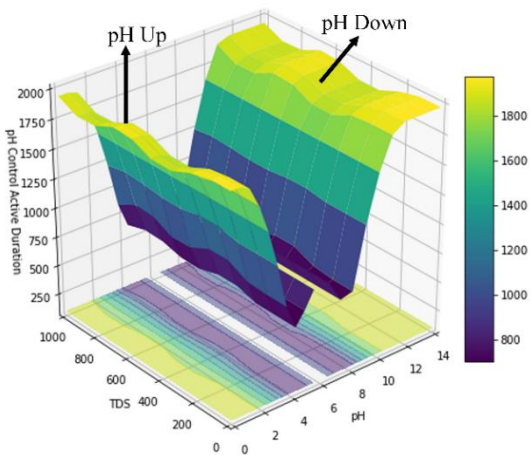

Fig. 7. pH surface control

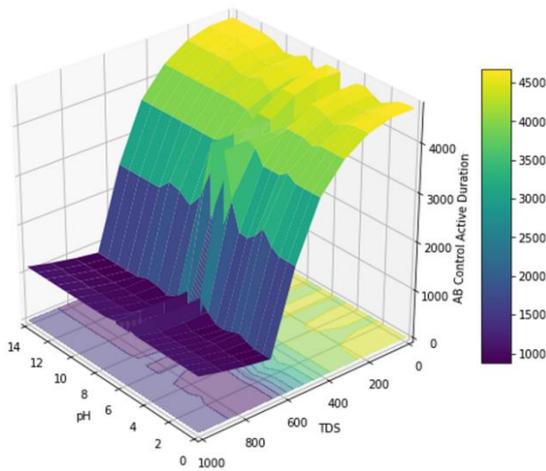

Fig. 8. AB mix surface control

The difference between pH control and AB mix control is in the resulting duration range. pH control appears to be symmetrical for pH Up and pH Down, with a maximum duration of 2000 ms. Meanwhile, the AB mix has a maximum duration of about 4500 ms and begins to converge at TDS levels above 600 ppm.

*B. Real Experiment on NFT Hydroponics*

The NFT hydroponic rack that has been created is shown in Fig. 9, and the control panel device in Fig. 10. Table 2 displays the results of six experiments, each with a different initial pH and TDS level. The nutrient's initial state was adjusted using a combination of normal, acid, alkaline, and very low PPM states. The initial state is manually achieved by pouring the nutrients, pH Up, and pH Down solutions, and is electronically measured using sensors and a microcontroller. The experiment was repeated until the final reading of the nutrient solution returned to normal. A 60-second delay is applied to one control cycle to allow the solution to mix evenly in the nutrient reservoir.

In the 5th experiment, it can be seen that the initial reading level of pH is 10.55 and TDS is 324 ppm. The fuzzy system works in 3 steps (repetitions) to reach normal conditions with a total duration of 3691 milliseconds of the pH Down pump and 4261 milliseconds of the AB mix pump. Similar results occurred in the 3rd and 6th experiments, where the system worked in 2 steps, while in the 1st, 2nd, and 4th, the fuzzy system was able to control the nutrient solution until it reached a normal state in one step. When the pH and TDS levels are within normal parameters, the fuzzy system produces a defuzzification output of 0 milliseconds or the pump is not active.

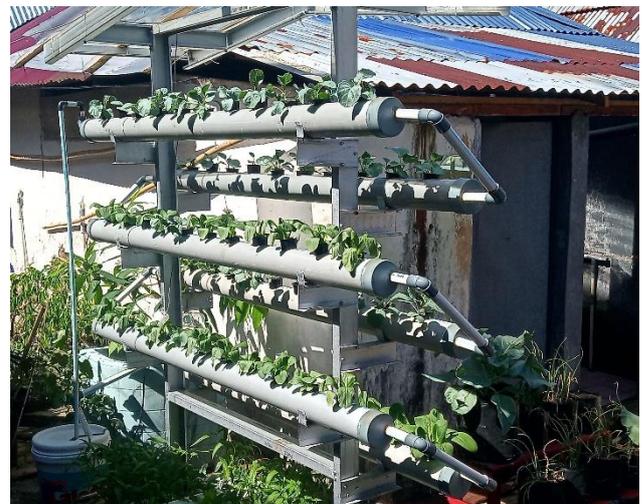

Fig. 9. NFT Hydroponic rack design results

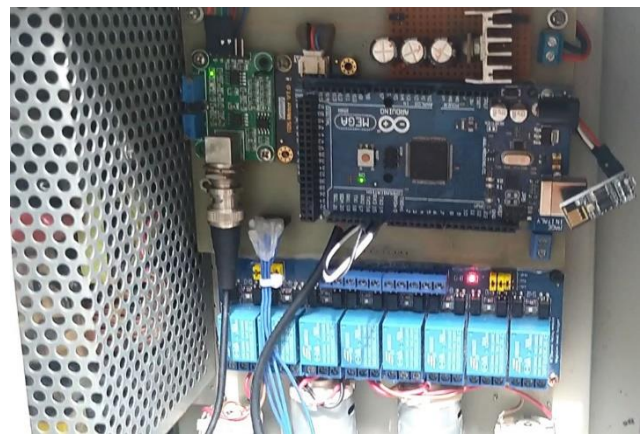

Fig. 10. Control panel circuit box

TABLE II. SAMPLES OF EXPERIMENTAL DATA FOR CONTROLLING pH AND TDS WITH FIS-MAMDANI

| Experiment | | 1 | 2 | 3 | | 4 | 5 | | | 6 | |
|---|---|---|---|---|---|---|---|---|---|---|---|
| Initial value | pH | 6.35 | 6.09 | 4.02 | 4.96 | 4.54 | 10.55 | 8.67 | 6.83 | 9.46 | 7.69 |
| | TDS (ppm) | 110 | 946 | 272 | | 117 | 324 | | | 531 | |
| pH Up active duration (ms) | | 0 | 0 | 663 | 773 | 689 | 0 | 0 | 0 | 0 | 0 |
| pH Down active duration (ms) | | 0 | 0 | 0 | 0 | 0 | 1786 | 1085 | 820 | 1481 | 651 |
| AB mix active duration (ms) | | 4274 | 1083 | 4310 | | 4370 | 4261 | | | 3212 | |
| Final value | pH | 6.34 | 6.11 | 4.96 | 6.12 | 5.67 | 8.67 | 6.83 | 5.72 | 7.69 | 6.45 |
| | TDS (ppm) | 1124 | 1247 | 1274 | | 1210 | 1255 | | | 1300 | |

The time response samples are shown in Fig. 11, Fig. 12, and Fig. 13. Each image shows the response time of the pH Up control in the 3rd experiment, pH Down in the 6th, and the TDS control in the 1st experiment. The blue line curve on the pH Up control time response in Fig. 11 is the first step response, and because the normal pH of 5.5 - 6.5 has not been reached, the process continues to second step, which is the red line curve. The active duration of the pH pump may not be enough to raise the pH due to the water circulation and the AB mix nutrient solution that can decrease the pH level. So that the 3rd experiment required two control steps.

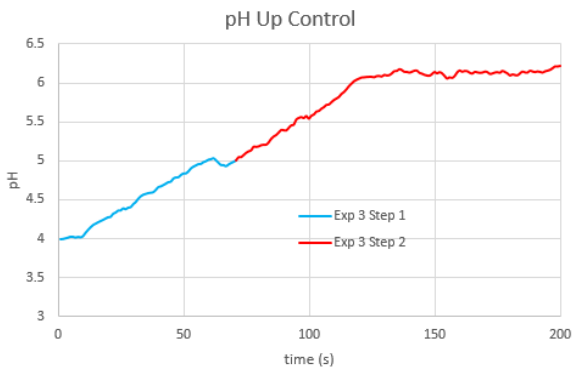

Fig. 11. pH Up time response

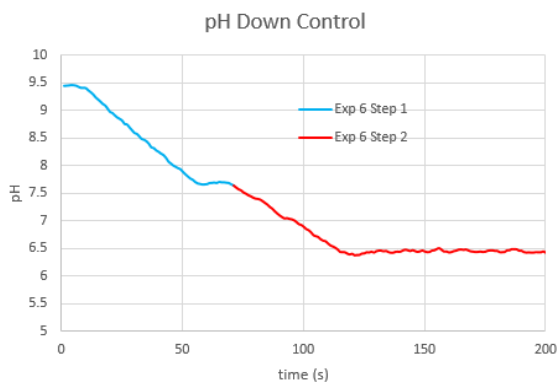

Fig. 12. pH Down time response

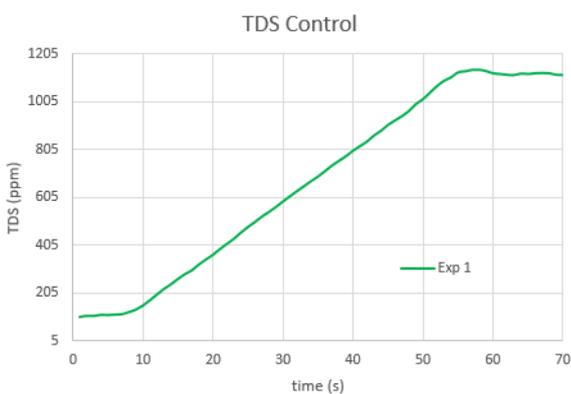

Fig. 13. TDS time response

In the 5th and 6th experiments, the initial pH level was high, the control system reduced the pH level by more than one step, usually due to the water circulation factor. If the fertilizer pump was not activated in the experiments to normalize TDS, the number of steps for normalizing pH could be increased. To be able to do control in one step, the pH control system can be optimized again by rearranging the degree of membership of the fuzzy set. By taking into account the possibility of pH Up and pH Down oscillations if targeted in one step, the current system is safer to use. As for TDS control, all experiments only require one step. The active duration of the fertilizer pump can accurately regulate TDS levels.

### C. Comparison of Real Experiment Versus Simulation

The real experimental results were compared with the simulation to validate the FIS Mamdani design used in NFT Hydroponic. Figs. 14, 15, and 16 show visual output from the FIS Mamdani controller simulation for pH Up, pH Down, and AB mix active duration. The 4th experiment shows the pH Up active duration of 689.09 ms for the input pH 4.54 and TDS 272 ppm, while AB mix active duration is 4233.24 ms. The 5th experiment step1 shows the pH Down active duration 1800.89 for the input pH 10.55 and TDS 324 ppm. There were some differences between the real experiment and the simulation. Figs. 17 and 18 show the outcome of the comparison.

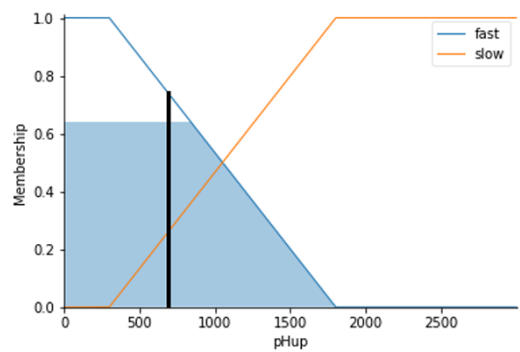

Fig. 14. 4th experiment: pH Up pump active duration = 689.09 ms.

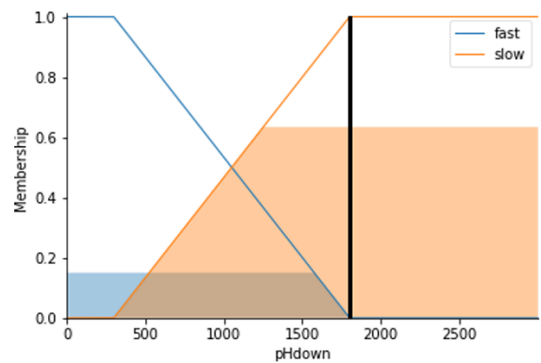

Fig. 15. 5th experiment step 1: pH Down pump active duration = 1800.89ms

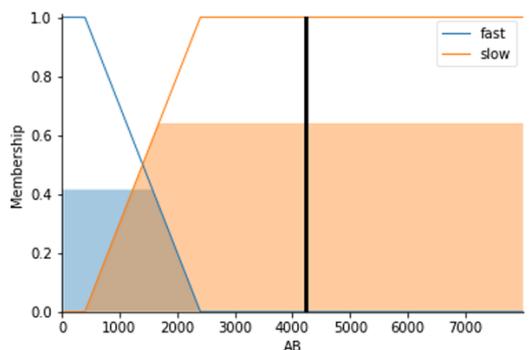

Fig. 16. 4th experiment: AB mix pump active duration = 4223.34 ms.

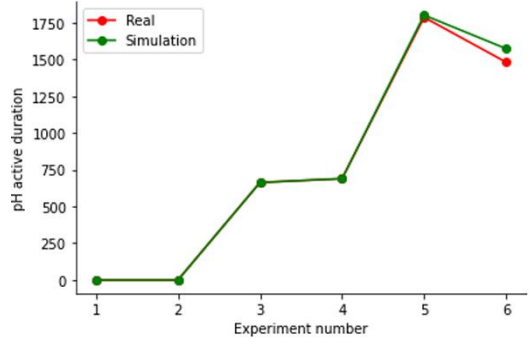

Fig. 17. Error of pH active duration of real experiment vs. simulation, rmse = 224.42 ms, normalized rmse = 0.4083, error max = 91.86 ms.

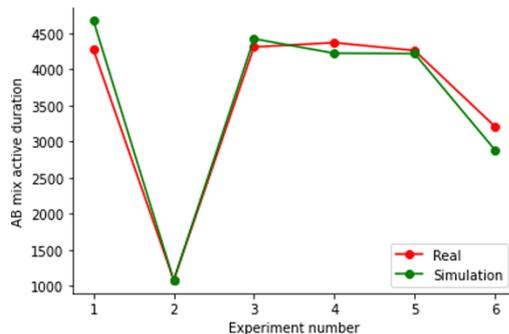

Fig. 18. Error of AB mix active duration of real experiment vs. simulation, RMSE = 224.42 ms, normalized RMSE = 0.4083, error max = 400.69 ms.

The maximum error in pH active duration is 91.86 ms, the RMSE is 224 ms, and the normalized RMSE is 0.4083. Meanwhile, the maximum error of AB mix active duration is 400.69 ms, with an RMSE of 224 ms and a normalized RMSE of 0.4083, which is the same as normalizing RMSE at pH active duration. This error data demonstrates that the microcontroller-based calculation of pH and AB mix active duration of FIS Mamdani can function properly. The fact that the FIS Mamdani design used on the microcontroller has the same normalized RMSE value indicates that it can operate consistently.

*D. IoT interface*

The EPS01 module and the Blynk IoT application were used to create an IoT monitoring system. The application displays data such as volume, pH levels, and TDS nutrient levels. Fig. 19 shows the display of the successfully designed monitoring application. The microcontroller is programmed to send data every 5 minutes. The speed of data transmission is determined by the internet network being used.

## IV. CONCLUSION

In this research, the FIS-Mamdani control system has been designed to control pH levels and TDS levels of NFT hydroponic nutrients. The system is also equipped with an IoT monitoring system. The input of the control system is the pH and TDS levels from the sensor, processed with FIS Mamdani, and produces the output of the active time duration of the pH Up pump, pH Down pump, and AB mix pump. The surface control, the simulation, and the real experiment data show that the control system is running well; it can normalize pH and TDS according to the desired normal standard. In the pH control system, there can be more than one step in the control process. This is possible due to the circulation and nutrition factors of the AB mix, but this can avoid the oscillations of the pH Up and pH Down pumps, so it is safer to use. And for TDS control, FIS-Mamdani can accurately predict the active time duration of the AB mix pump in one step. The NFT hydroponic IoT monitoring system is designed using the IoT Cloud Blynk application and displays volume status, pH levels, and TDS levels with a 5-minute delay.

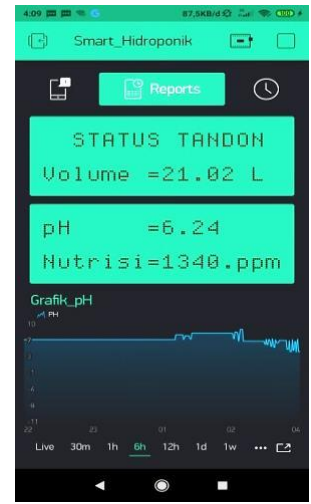

Fig. 19. IoT monitoring application display on smartphone